\documentclass{webofc}
\usepackage[varg]{txfonts}   % Web of Conferences font
\usepackage{graphicx}

\begin{document}

\title{Exploring Photoproduction with the GlueX Experiment}

\author{\firstname{Matthew} \lastname{Shepherd}\inst{1}\fnsep\thanks{\email{mashephe@indiana.edu}}}
        % etc.
\institute{Indiana University, Bloomington, IN 47405, USA}

%% PAGE LIMIT IS 6 PAGES FOR A PLENARY TALK

\abstract{
The GlueX experiment at Jefferson Lab (Newport News, VA USA) is designed
to explore the spectrum of mesons up to about 3 GeV.  We present 
results on the production of light-quark resonances with linearly polarized 
photons.  These results enhance our understanding of 
photoproduction mechanisms, which is valuable in subsequent searches
for exotic hybrid mesons.  Measurements of the $J/\psi$ photoproduction
cross section at threshold are also presented.
}

\maketitle

\section{Introduction}
\label{intro}

An outstanding puzzle in our study of strong interactions is understanding how 
the spectrum of mesons arises from Quantum Chromodynamics.  There is a growing 
body of evidence that QCD generates mesons and baryons beyond quark-antiquark 
and three-quark configurations.  In addition, theoretical calculations of 
the spectrum of mesons predict states with gluonic degrees of freedom that arise 
from the gluon-gluon interaction in QCD.  A key objective of the GlueX 
experiment is search for these hybrid mesons in the light quark sector 
using linearly polarized photon beams, which provide sensitivity to 
production dynamics.  We present a collection of initial results that are
useful for developing theoretical and experimental tools for the study of
meson photoproduction and ultimately the search for hybrid mesons.

\section{The GlueX Experiment}
\label{gluex}

The GlueX experiment is designed to explore both production and decay of hadronic
resonances.  The linearly polarized photon beam, which provides access to experimental
observables related to production mechanisms, is derived through coherent bremsstrahlung
radiation of the 12 GeV electron beam from Jefferson Lab's Continuous Electron Beam
Accelerator Facility (CEBAF) on a thin diamond radiator.  The linearly polarized
photon beam impinges on a liquid hydrogen (proton) target and the reaction products
are detected by the GlueX detector, which has a 
large acceptance for both charged particles and photons.  All results presented in
these proceedings rely on exclusive reconstruction of the full reaction to suppress
backgrounds.  A detailed discussion of the apparatus can be found in 
Ref.~\cite{gluex_nim}.

The photon beam, whose energy and polarization is independently measured with
beamline instrumentation, has a peak polarization of about 35\% in the region
$8.2 < E_\gamma~[\mathrm{GeV}] < 8.8$.  If one considers only the portion of
the beam from that energy region, the photon flux on target peaked at about 
$5\times 10^7~\gamma$/s and the total integrated luminosity on a proton
target collected by GlueX through calendar year 2020 is about 250~pb$^{-1}$.  
For comparison, the $\rho$ meson cross section is about $10~\mu$b, and 
resonances like the $a_2$ have cross sections around tens to hundreds of 
nanobarns.

%%%%%

\section{Spin-Density Matrix Elements for $\rho$ Meson Production}

\begin{figure}
\centering
\sidecaption
\includegraphics[width=0.7\linewidth]{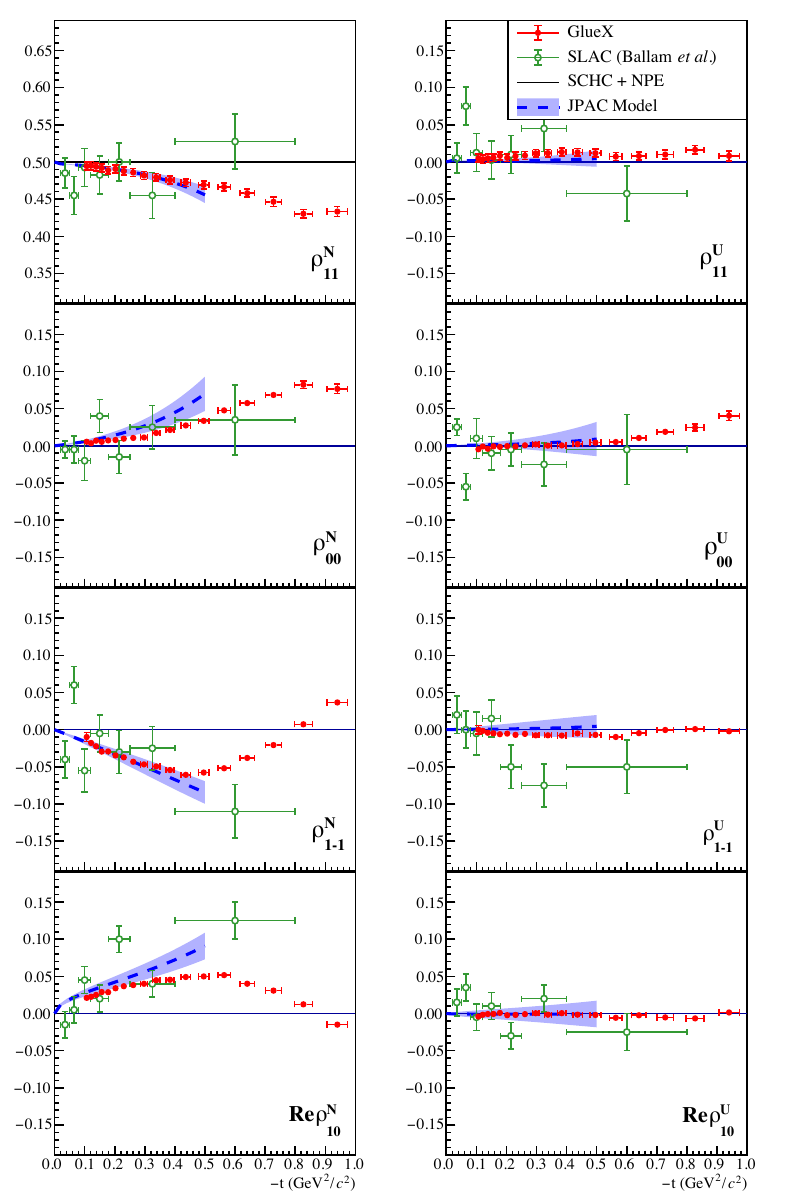}
\caption{The spin-density matrix elements for $\rho$ meson production 
are separated 
into natural parity exchange (left column) and unnatural parity exchange 
(right column).  The existing data from SLAC~\cite{ballam} are shown along 
with the GlueX results.  The data are all compared with a model calculation 
by the Joint Physics Analysis Center (JPAC)~\cite{jpac_rho}.  
The limit of pure $s$-channel helicity conservation and natural parity 
exchange is given by the horizontal black lines.}
\label{fig:rho_sdme}
\end{figure}

An early goal of the GlueX experiment is to develop an understanding of how
to model the photoproduction of meson systems.  At beam energies relevant
for the GlueX environment, the production of mesons is often modelled as
a $t$-channel exchange of a virtual particle (a ``Reggeon") with the target.
A feature of using linearly polarized photon beams to produce mesons is
that it allows access to the naturality of the Reggeon.  (Natural parity
is defined $P = (-1)^J$, whereas unnatural parity is the opposite.)  In
the production of spinless mesons, the angular distribution of the production
plane with respect to linear polarization can be used to extract what
is known as the beam asymmetry ($\Sigma$).  Early results from GlueX
showed the dominance of natural parity exchange~\cite{eta_ba}, 
except at low momentum transfer ($t$) when pion exchange 
is allowed~\cite{pim_ba}.

For the production of a vector resonance like the $\rho^0,$ which decays
to $\pi^+\pi^-$,  the production and polarization of the $\rho$ can be 
characterized by a set of spin-density matrix elements (SDMEs) $\rho(\rho^0)^k_{ij}$,
which are connected to the initial photon spin-density matrix $\rho(\gamma)$ by
a transition amplitude $T$:  $\rho(\rho^0) = T\rho(\gamma)T^*$~\cite{schilling}.  
These SDMEs convey not only the quantity of natural and unnatural exchange with
the target, but also the transfer of polarization from the incoming
polarized photon to the outgoing $\rho$ resonance.  At very low momentum
transfer, $\rho$ production should proceed through $s$-channel helicity
conserving natural parity exchange (SCHC-NPE), {\it i.e.,} the $\rho$ exchanges
a particle that has vacuum quantum numbers with the target and emerges
with same polarization as the incident photon.  As $|t|$ increases the
interaction is expected to deviate from this simplistic picture.

The SDMEs are measured in each region of $|t|$ by performing a
fit of three-dimensional intensity function to the data, where the
three relevant angles are the angle that describes the $\rho$ production
with respect to the beam polarization and the two angles that describe
the decay of the $\rho$ and hence encode the $\rho$ polarization.  
In the high energy limit, linear combinations of the SDMEs can be used
to isolate the natural and unnatural parity exchange components~\cite{schilling}, 
specifically $\rho^{\mathrm{N,U}}_{ik}=\left(\rho^0_{ik}\mp (-1)^i\rho^1_{-ik}\right)/2$.
The evolution of these SDMEs as a function of $|t|$ is shown in Fig.~\ref{fig:rho_sdme}.
For all values of $t$ the natural parity exchange process dominates over
unnatural parity exchange.  The deviation from SCHC ($\rho^{N}_{11}=\frac{1}{2})$ 
is due to natural parity exchange amplitudes that increase in 
magnitude with momentum transfer.  See Ref.~\cite{rhosdme} for additional 
details.

\section{The Search for Hybrid Mesons}

The gluon self-interaction in QCD naively should permit construction of 
hybrid mesons with gluonic degrees of freedom: $q\bar{q}g$.  Calculations
of the spectrum of light mesons in QCD~\cite{lqcd_spectrum} reveal a 
rich pattern of states beyond the traditional $q\bar{q}$ states that
have an operator overlap consistent with a hybrid meson interpretation.
A striking experimental signature of hybrids is that some of them have
exotic $J^{PC}$.  To date, the best evidence for the existence of
hybrid mesons comes from an analysis of COMPASS data on pion production
of $\eta\pi$ and $\eta'\pi$~\cite{compass_pi1,jpac_pi1}.  (This work 
was subsequently extended to include data from Crystal 
Barrel~\cite{cb_pi1}.)  The data are consistent with a broad isovector 
exotic $J^{PC}=1^{-+}$ resonance denoted $\pi_1(1600)$ that couples to 
both $\eta\pi$ and $\eta'\pi$.  While the evidence for this state is
strong, one really desires the identification of multiple members of a
spectrum of hybrid mesons to understand the role of gluonic degrees of freedom
in the hadron spectrum.  Further progress towards this goal was made by
the BESIII Collaboration who observed a candidate for the exotic $\eta_1$ or
$\eta'_1$~\cite{besiii_eta1}.  An initial target of the GlueX 
Collaboration is to use photoproduction to confirm the existence of 
the $\pi_1(1600)$ and, if observed, study its production mechanisms.

\subsection{Upper Limits on the $\pi_1(1600)$ Photoproduction Cross Section} 

\begin{figure}
\centering
\sidecaption
\includegraphics[width=0.7\linewidth]{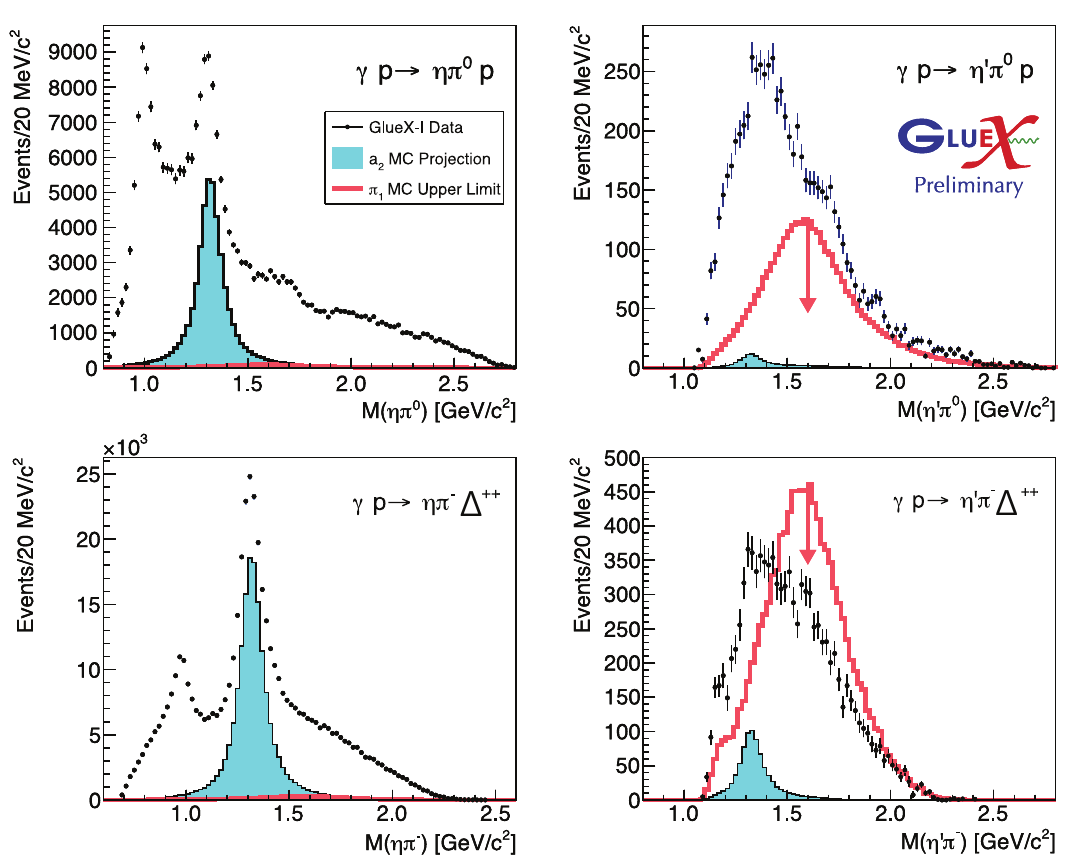}
\caption{Invariant mass spectra for the $\eta\pi$ (left) and $\eta'\pi$ 
(right) systems produced in neutral-exchange recoil proton reaction 
(upper) and charge-exchange recoil $\Delta^{++}$ reaction (lower). 
The projected size of the $a_2^{-,0}$ signal is shown, using branching 
ratios from the PDG.  The red curves show upper limits on the 
contribution of exotic $\pi_1$ to the spectra.}
\label{fig:pi1_limit}
\end{figure}

In contrast to spinless pion beams, the beam photon can be pictured
as a spin-one virtual vector meson.  The alignment of quark spins in
such states was suggested by some models as a reason for enhanced
exotic hybrid cross sections in photoproduction -- the $\pi_1(1600)$
is thought to be the exotic member of a triplet of hybrid states 
($0^{-+}$, $1^{-+}$, and $2^{-+}$) all with $S_{q\bar{q}} = 1$.  
A challenge with exploring the hypothesis of enhanced photoproduction
cross sections of hybrids is that one is experimentally sensitive to 
the product of production and decay couplings of states.  Therefore, it 
is often difficult to learn about the production cross section from
a single measurement.  In recent years, calculations of partial 
widths of exotic hybrid resonances have been performed using 
Lattice QCD~\cite{lqcd_pi1_width}.  According to these calculations the
dominant decay mode of the $\pi_1(1600)$ is $b_1\pi$.  The presence
of a single dominant decay mode in the calculation provides a
lower limit for the decay coupling and hence the ability to set an
upper limit on the production cross section of $\pi_1(1600)$ using
data from GlueX.  Most importantly, this upper limit, coupled again
with Lattice QCD calculations of branching ratios, can be used to
guide the search for exotic hybrid mesons to the most sensitive
reactions.

To obtain upper limits on $\sigma(\gamma p \to \pi_1^- \Delta^{++})$ 
and $\sigma(\gamma p \to \pi_1^0 p)$ we first note that the
predicted dominant $b_1\pi$ decay mode of the $\pi_1$ will produce the final state
$\omega\pi\pi$.  Using GlueX data, we reconstruct the reactions $\gamma p \to \omega\pi^+\pi^- p$
and $\gamma p \to \omega\pi^0\pi^0 p$ and use isospin relations to isolate
the cross section for $\gamma p \to (\omega\pi\pi)_{I=1} p$ where we might
expect to observe the $\pi_1^0$ as a peak in the $\omega\pi\pi$ mass spectrum.
In addition, we analyze the isospin-one cross section 
$\gamma p \to (\omega\pi\pi)^- \Delta^{++}$.  No distinct $\pi_1$ peak appears
in either spectrum but assuming the $\pi_1$ saturates the spectra provides
an upper limit for the cross section.  Quantitatively, the upper limit
for $\pi_1^-$ photoproduction cross section is at the level of the
$a_2(1320)^-$ cross section.  A similar result is obtained for the $\pi_1^0$ 
compared to the $a_2(1320)^0$ cross section.  These results suggest the exotic $\pi_1$
photoproduction cross section is not significantly enhanced with respect to 
the conventional $a_2$ meson.

Equally as important, the results, again coupled with LQCD calculations
of branching ratios, allow one to set limits on the contribution of 
the $\pi_1$ to various spectra.  These {\em upper limits} are shown by the
red curves in Fig.~\ref{fig:pi1_limit}.  Our results exclude a large
contribution of the exotic amplitude to the $\eta\pi$ mass spectrum. 
However, a large contribution to the $\eta'\pi^{-,0}$ mass spectra is
not excluded, and consequently these reactions provide the best sensitivity
in a search for the $\pi_1(1600)$ in photoproduction.  Using measured
branching ratios from the Particle Data Group, we can also predict the
$a_2$ contribution to each spectrum.  Enhancements at the $a_2$ mass
in the data are evident in all spectra.  An understanding of conventional
$a_2$ production and polarization is essential as it will act as both
a standard candle and interferometer in the search for the exotic $\pi_1$.

\subsection{Photoproduction of $a_2(1320)$}

\begin{figure}
\centering
\sidecaption
\includegraphics[width=0.5\linewidth]{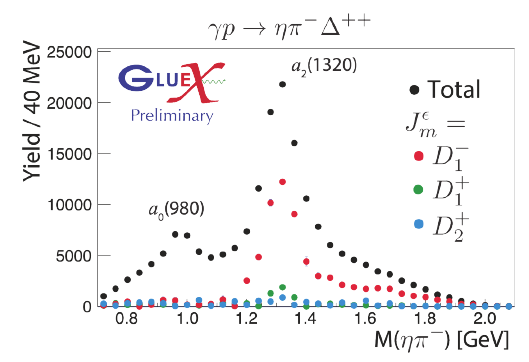}
\caption{ The $\eta\pi^-$ mass spectrum (black points) in the reaction 
$\gamma p \to \eta \pi^- \Delta^{++}$ for $0.1 < |t|~[\mathrm{GeV}] < 0.3$.  
The colored points show the result of an amplitude analysis in 
bins of $\eta\pi^-$, where the amplitudes for spin $J$, spin-projection 
$m$, and reflectivity $\epsilon$ are denoted by $J_m^\epsilon$. }
\label{fig:a2_minus}
\end{figure}

As shown in Fig.~\ref{fig:pi1_limit}, the $a_2(1320)$ is clearly visible as
a peak in both the $\eta\pi^-$ and $\eta\pi^0$ spectra.  In order to
facilitate the search for exotic mesons in $\eta'\pi$, where the
$a_2$ is comparatively smaller, we need to use the $\eta\pi$ channels
to constrain our knowledge of the $a_2$.  In the case of 
$\gamma p \to a_2^0p$ there are ten different amplitudes that provide
a description of the production:  $(2J+1)$ possible projections of the
$a_2$ spin times two choices of reflectivity $\epsilon$, which, in the high-energy
limit, corresponds to production by natural ($\epsilon=+$) and unnatural 
($\epsilon = -$) exchange~\cite{mathieu_etapi}.  At low momentum transfer
to the target, the $a_2^0$ is produced dominantly by natural parity exchange
in an $m=2$ state denoted $D_2^+$ (see Ref.~\cite{ma_hadron} for details).  Interestingly,
a similar helicity structure is also observed in two-photon production 
of the $a_2$~\cite{a2_2gamma}.  Figure~\ref{fig:a2_minus} shows the dominant
$D$-wave amplitudes in the $\eta\pi^-$ system at low momentum transfer to the
target.  In addition to the $a_2(1320)^-$ additional intensity in
the $D$-wave at higher mass, consistent with the $a_2'$, is observed.
In contrast to the $a_2^0$, the $a_2^-$ is produced dominantly 
in the $D_1^-$ amplitude, which is
consistent with a pion exchange process where the $a_2$ emerges with
a polarization that matches the incident beam.

\section{Photoproduction of $J/\psi$ at Threshold}

Measuring the total cross section ($\sigma$) and differential cross section
($d\sigma/dt$) of $\gamma p \to J/\psi p$ as a function of beam energy $E_\gamma$
provides access to a variety of topics.  One model for the production
process involves conversion of the photon to the $c\bar{c}$ system via the exchange
of high-momentum gluons with the proton.  In this picture, the dependence
of the cross section on momentum transfer to the proton gives access to 
the gluonic form-factors of the proton~\cite{jpsi_gg_ff}.  On the topic
of hadron spectroscopy, it is interesting to look for peaks in the cross
section as a function of energy.  The LHCb collaboration has reported several
candidates for pentaquark resonances that couple to $J/\psi p$~\cite{pc_lhcb}.  
Naively, one would expect to excite such resonances in photon-proton collisions
at of the appropriate energy.

\begin{figure}
\centering
\sidecaption
\includegraphics[width=0.6\linewidth]{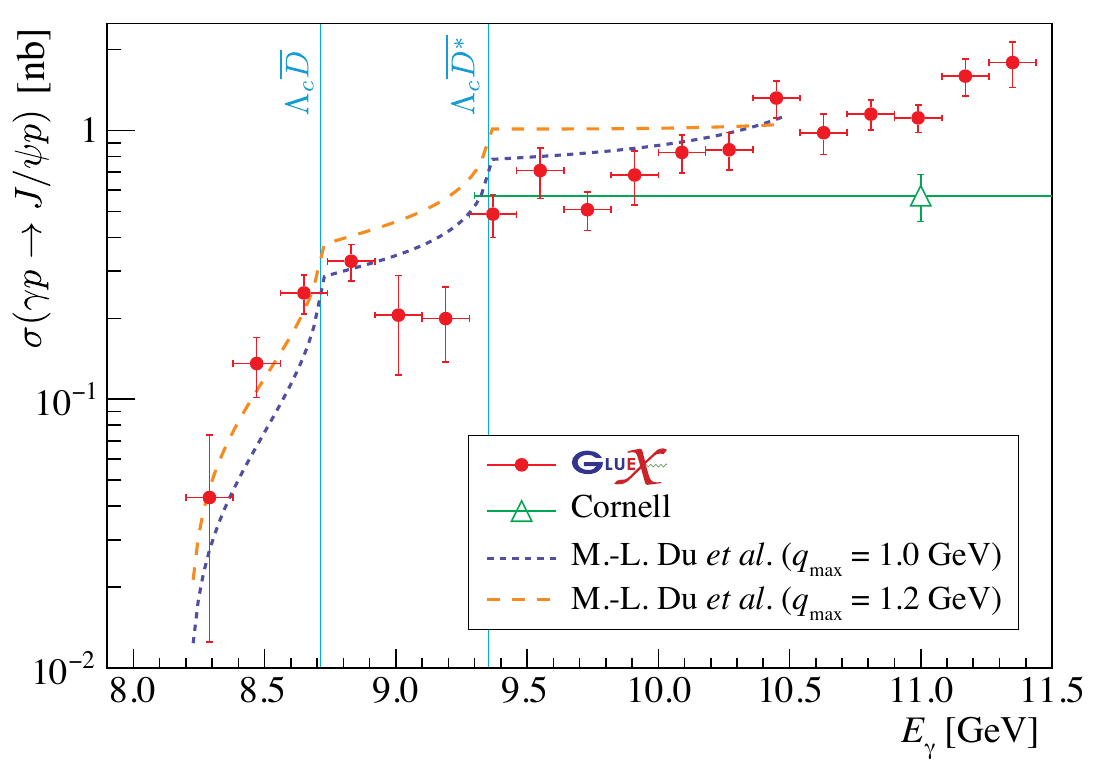}
\caption{The $J/\psi$ photoproduction cross section as a function 
of photon beam energy measured by GlueX (red points).  A previous 
measurement from Cornell~\cite{jpsi_cornell} is shown in the 
green point.  The dashed and dotted lines show cross section 
calculation for a model that includes the effects of open charm 
thresholds $\Lambda_c\overline{D}$ and 
$\Lambda_c\overline{D^*}$~\cite{mldu_jpsi}.}
\label{fig:jpsi}
\end{figure}

Experimentally, the $J/\psi$ is detected through its $e^+e^-$ decay mode,
which leaves a unique signature of two high-energy electromagnetic showers
in the detector.  The recent GlueX measurements~\cite{jpsi_prc} for 
the production cross section of $J/\psi$ 
as a function of beam energy are shown in Fig.~\ref{fig:jpsi}.
These measurements reveal an interesting cusp-like structure in the 
energy dependence of the cross section that coincides with thresholds
for production of $\Lambda_c\overline{D}$ and $\Lambda_c\overline{D^*}$.
Additional theoretical work and more precise measurements, especially
those with polarization observables, are needed to understand nature
of these structures, which may be evidence of resonances or other effects
that have implications on the ability to interpret the data in the context
of proton gluonic form factors~\cite{jpac_jpsi}.

\section{Summary and Acknowledgements}

In summary, a collection of results from the initial running of the
GlueX experiment have been presented.  Measurements of $\rho$ spin-density
matrix elements and the properties of $a_2$ production aid in refinement 
of photoproduction models.  Upper limits on  $\pi_1$ photoproduction
cross sections coupled with Lattice QCD results guide the search for
exotic mesons.  A study of $J/\psi$ production provides an opportunity
to enhance our understanding of the proton and constrain properties of
candidates for pentaquarks containing $c\bar{c}$.

I would like to thank my colleagues in the GlueX Collaboration and
the MESON 2023 organizers for the opportunity to present these results.
This work was funded in part by the US Department of Energy Office of
Nuclear Physics under grant DE-FG02-05ER41374.

% BibTeX or Biber users please use (the style is already called 
% in the class, ensure that the "woc.bst" style is in your local directory)
% \bibliography{name or your bibliography database}
%
% Non-BibTeX users please use
%

\end{document}